\documentclass[3p,times,procedia]{elsarticle}
\usepackage{nupha_ecrc}
\usepackage{graphicx}

\usepackage[    bookmarks,
                 bookmarksopen = true,
                 bookmarksnumbered = true,
                 linktocpage,
                 colorlinks = true,
                 linkcolor = blue,
                 urlcolor  = blue,
                 citecolor = purple,
                 anchorcolor = green,
                 hyperindex = true,
                 hyperfigures]
                 {hyperref}

\volume{00}

\firstpage{1}

\journalname{Nuclear Physics A}

\runauth{L-G. Pang et al.}

\jid{nupha}

\jnltitlelogo{Nuclear Physics A}

\usepackage{amssymb}

\biboptions{comma,square,compress}

\usepackage[figuresright]{rotating}

\begin{document}

\begin{frontmatter}



\dochead{XXVIth International Conference on Ultrarelativistic Nucleus-Nucleus Collisions\\ (Quark Matter 2017)}

\title{Local and global $\Lambda$ polarization in a vortical fluid}

\author[ustc]{Hui Li}
\author[fias,itp,gsi]{Hannah Petersen}
\author[fias,ucb,lbnl]{Long-Gang Pang* \corref{cor1}}
\ead{lgpang@lbl.gov}
\author[ustc]{Qun Wang}
\author[ustc]{Xiao-Liang Xia}
\author[ccnu,lbnl]{Xin-Nian Wang}

\address[ustc]{Department of Modern Physics, University of Science and Technology of China, HeFei, 230026, China}
\address[fias]{Frankfurt Institute for Advanced Studies, Frankfurt am Main, 60438, Germany}
\address[itp]{Institute for Theoretical Physics, Goethe University, Frankfurt, 60438, Germany}
\address[gsi]{GSI Helmholtzzentrum f\"ur Schwerionenforschung, Darmstadt, 64291, Germany}
\address[ucb]{Physics Department, University of California, Berkeley, CA 94720, USA}
\address[lbnl]{Nuclear Science Division, Lawrence Berkeley National Laboratory, Berkeley, CA 94720, USA}
\address[ccnu]{Key Laboratory of Quark and Lepton Physics (MOE) and Institute of Particle Physics, Central China Normal University, Wuhan, 430079, China}

\begin{abstract}

We compute the fermion spin distribution in the vortical fluid created in off-central high energy heavy-ion collisions. We employ the event-by-event (3+1)D viscous hydrodynamic model. The spin polarization density is proportional to the local fluid vorticity in quantum kinetic theory. As a result of strong collectivity, the spatial distribution of the local vorticity on the freeze-out hyper-surface strongly correlates to the rapidity and azimuthal angle distribution of fermion spins. We investigate the sensitivity of the local polarization to the initial fluid velocity in the hydrodynamic model and compute the global polarization of $\Lambda$ hyperons by the AMPT model. The energy dependence of the global polarization agrees with the STAR data.

\end{abstract}


\end{frontmatter}


\section{Introduction}
\label{sec:intro}

Recently the STAR collaboration has measured the polarization of $\Lambda$ and $\bar{\Lambda}$ hyperons \cite{Nocera:2017wep}. They observed that (a) the global polarization decreases with collision energies; (b) the polarization for $\bar{\Lambda}$ is always bigger than $\Lambda$ at the same beam energy. These two features are very important to quantitatively constrain the fluid vorticity of the strongly coupled quark-gluon plasma (sQGP) and the magnitude of the magnetic field through the spin-vorticity and spin-magnetic coupling \cite{Becattini:2016gvu,Shuryak:2016hor,Pang:2017bjc}. The measured beam energy dependence is consistent with the predictions of the hydrodynamic or transport model  \cite{Becattini:2016gvu,Liang:2004ph,Csernai:2013bqa,Becattini:2015ska,Jiang:2016woz,Deng:2016gyh,Pang:2016igs}. The difference between $\Lambda$ and $\bar{\Lambda}$ is caused by (1) pauli-blocking -- it is more difficult to polarize $\Lambda$s than $\bar{\Lambda}$s when there are more fermions than anti-fermions \cite{Fang:2016vpj}; (2) the spin-magnetic coupling -- it generates opposite contributions to $\Lambda$ and $\bar{\Lambda}$~\cite{Becattini:2016gvu,Shuryak:2016hor,Pang:2017bjc}. This difference provides a unique opportunity to determine the magnetic field created in heavy-ion collisions. In this note, we investigate the effect of the initial flow on the local $\Lambda$ polarization within a (3+1)D viscous hydrodynamic model and the beam energy dependence of the global $\Lambda$ polarization within A Multi-Phase Transport (AMPT)~\cite{Lin:2004en} model. This study deepens our understanding of the most vortical fluid ever produced~\cite{Nocera:2017wep}.
 
\section{$\Lambda$ polarization from (3+1)D viscous fluid dynamics}
\label{sec:model}
We employ the CLVisc model \cite{Pang:2016igs,Pang:2014ipa} to solve the energy-momentum conservation equation together with extended second-order Israel-Stewart equations for the shear viscosity, 
\begin{equation}
\nabla_{\mu} T^{\mu \nu} = 0; \;\;\; \Delta^{\mu\nu\alpha\beta} u^{\lambda} \nabla_{\lambda} \pi_{\alpha\beta} =  - \frac{1}{\tau_{\pi}}(\pi^{\mu \nu} -  \eta \sigma^{\mu\nu}) - \frac{4}{3}\pi^{\mu\nu} \theta + 2\pi^{\langle \mu}_{\lambda} \omega^{\nu \rangle \lambda} - \frac{\lambda_3}{\tau_{\pi}} \omega^{\langle \mu}_{\lambda} \omega^{\nu \rangle \lambda}
\end{equation}
with $T^{\mu\nu} \equiv \varepsilon u^{\mu}u^{\nu} - P \Delta^{\mu \nu} + \pi^{\mu \nu}$ being the energy-momentum tensor, $\varepsilon$ the energy density, $ u^{\mu}$ the fluid four-velocity normalized as $u_{\mu}u^{\mu}=1$, $P$ the pressure, $\Delta^{\mu \nu} \equiv g^{\mu\nu} - u^{\mu}u^{\nu}$ the transverse projector obeying $u_{\mu} \Delta^{\mu\nu}=0$. The $\eta$ is the shear viscosity, $\tau_{\pi}$ is the relaxation time, $\sigma^{\mu\nu} \equiv 2 \Delta^{\mu\nu\alpha\beta} \nabla_{\alpha} u_{\beta}$ is the symmetric shear tensor, $\omega^{\mu\nu} \equiv \Delta^{\mu\alpha}\Delta^{\nu\beta}(\nabla_{\alpha} u_{\beta} - \nabla_{\beta} u_{\alpha})$ is the anti-symmetric vorticity tensor and $\theta \equiv \nabla_{\mu} u^{\mu}$ is the expansion rate. $\Delta^{\mu\nu\alpha\beta} \equiv \frac{1}{2}(\Delta^{\mu\alpha}\Delta^{\nu\beta} + \Delta^{\mu\beta} \Delta^{\nu\alpha}) - \frac{1}{3}\Delta^{\mu\nu} \Delta^{\alpha\beta}$ is the double projector that renders the contracted tensor traceless and transverse to the fluid four-velocity. We observed that the coupling between $\pi^{\mu\nu}$ and $\omega^{\mu\nu}$ only introduces a small difference. The self-coupling term of $\omega^{\mu \nu}$ has an undetermined coefficient $\lambda_3$, which leads to a breakdown of the program for fluctuating initial conditions with $\lambda_3=1$. So these two terms are not included in the present work. The local polarization density on the freeze-out hyper-surface reads \cite{Fang:2016vpj,Becattini:2013fla},
\begin{equation}
  P^{\mu} \equiv \frac{d\Pi^\mu(p)/d^3 p}{dN/d^3p} = \frac{\hbar}{4m}\frac{\int d\Sigma_{\alpha} p^{\alpha} \Omega^{\mu\nu} p_{\nu} f(1 - f)}{\int d\Sigma_{\alpha} p^{\alpha} f},
   \label{eq:Pmu}
\end{equation}
where $m$ is the fermion mass, $d\Sigma_{\alpha}$ is the hyper-surface determined by the freeze-out  temperature $T_f=0.137$ GeV, and $\Omega^{\mu\nu} = \frac{1}{2} \epsilon^{\mu\nu\rho\sigma} \nabla_{\rho}\beta_{\sigma}$ is the thermal vorticity with $\beta_{\sigma}=u_{\sigma}/T$. Because of strong collectivity, the momentum distribution of the local polarization density is directly related to the local vorticity distribution in space-time \cite{Pang:2017bjc}.

\begin{figure}[htbp]
\begin{center}
\includegraphics[width=0.32\textwidth, trim={1.5cm 0.5cm 2cm 0.5cm},clip]{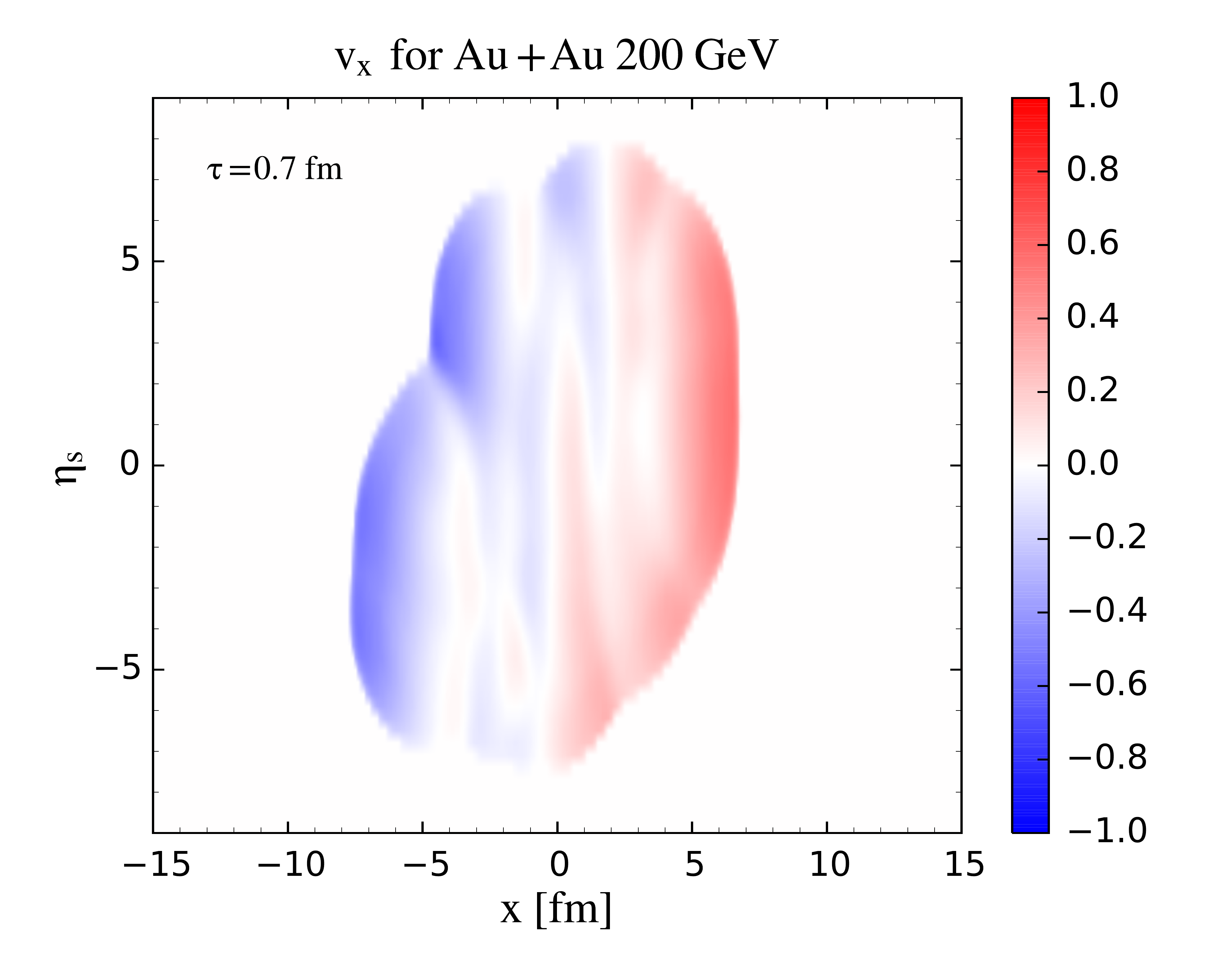}\hspace{0.1\textwidth}\includegraphics[width=0.32\textwidth, trim={1.5cm 0.5cm 2cm 0.5cm},clip]{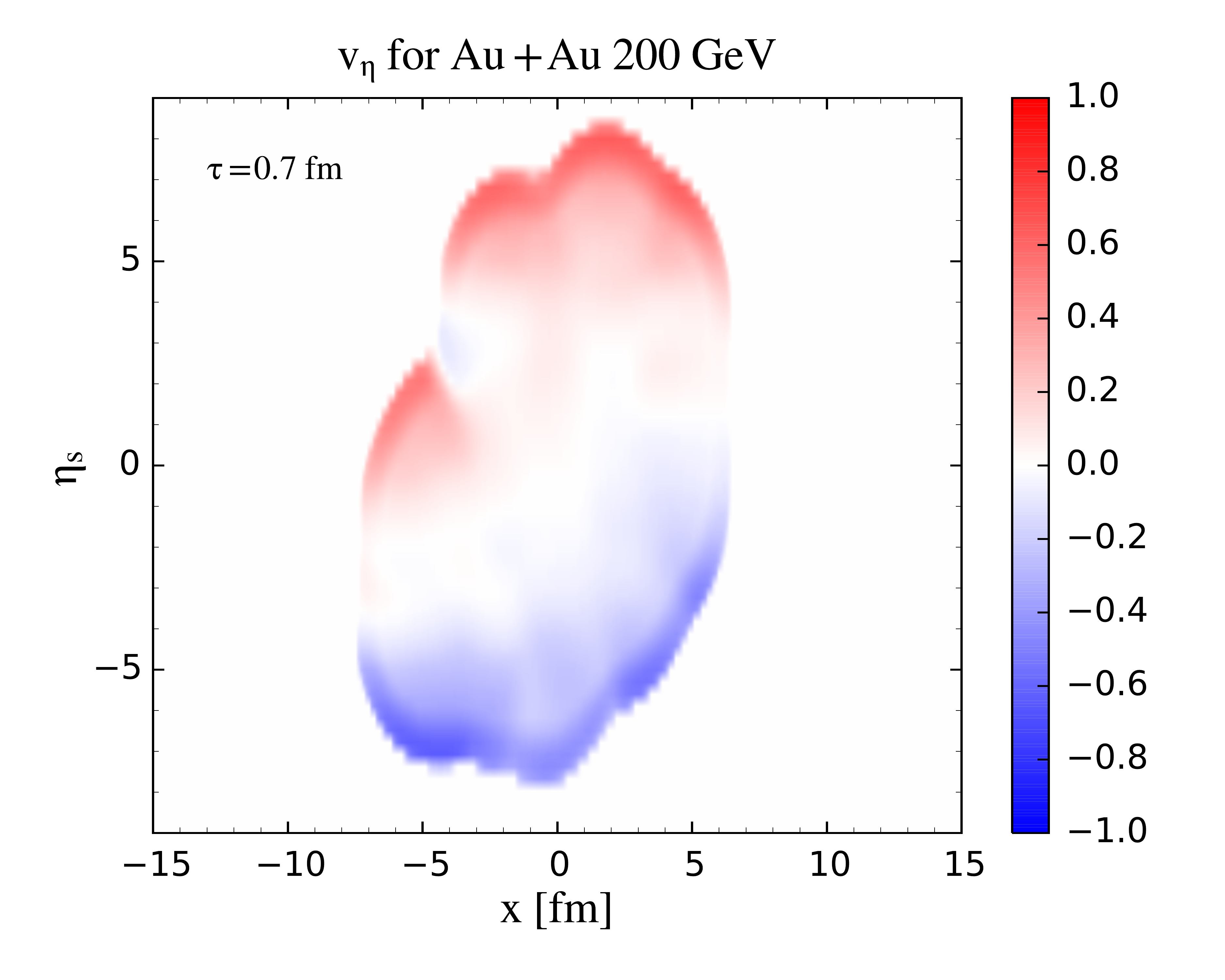}
\caption{The in-plane $v_x$ and $v_{\eta}$ distribution at $\tau=0.7$ fm after fluid dynamic evolution from fluctuating initial energy density distributions at $\tau_0=0.4$~fm from AMPT model.}
\label{fig:vx_vz}
\end{center}
\end{figure}

The AMPT model is employed to generate the initial energy density with the global angular momentum given by the asymmetry between forward and backward going participants separated by the impact parameter in the transverse plane. We start with the first assumption that at high beam energies there is only Bjorken flow at initial stage which leads to $v_x = v_y = v_{\eta} = 0$. In this case the deposited angular momenta are caused by the asymmetric distribution of the matter for $x>0$ (projectile side) and $x<0$ (target side), for one specific space-time rapidity. This asymmetry is shown in Fig.~\ref{fig:vx_vz}. The local vorticity is $0$ in the initial stage and generated during the fluid evolution.
\begin{figure}[htbp]
\begin{center}
\includegraphics[width=0.33\textwidth, trim={1.0cm 0.55cm 3cm 0.65cm},clip]{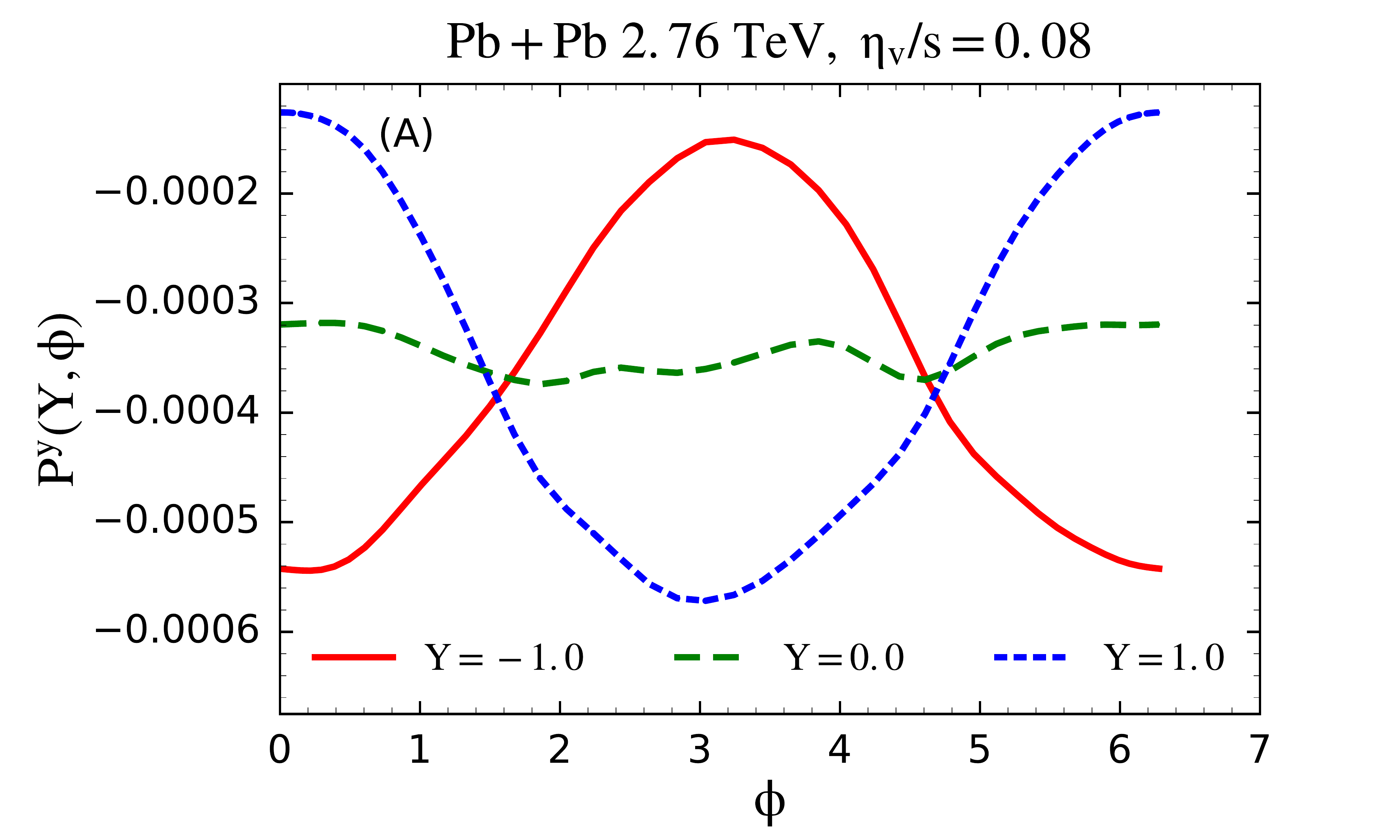}\includegraphics[width=0.33\textwidth, trim={1.0cm 0.55cm 3cm 0.65cm},clip]{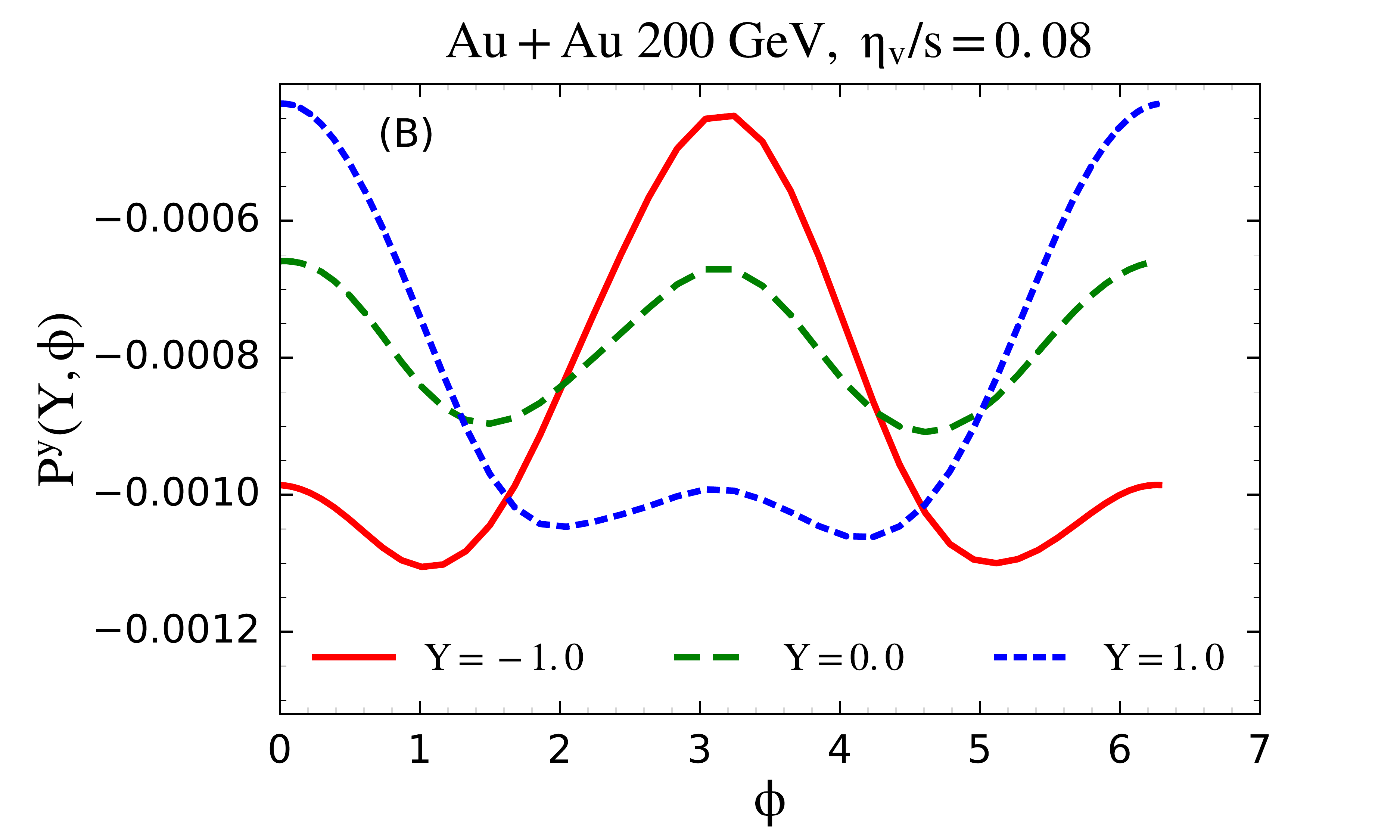}\includegraphics[width=0.33\textwidth, trim={1.0cm 0.55cm 3cm 0.65cm},clip]{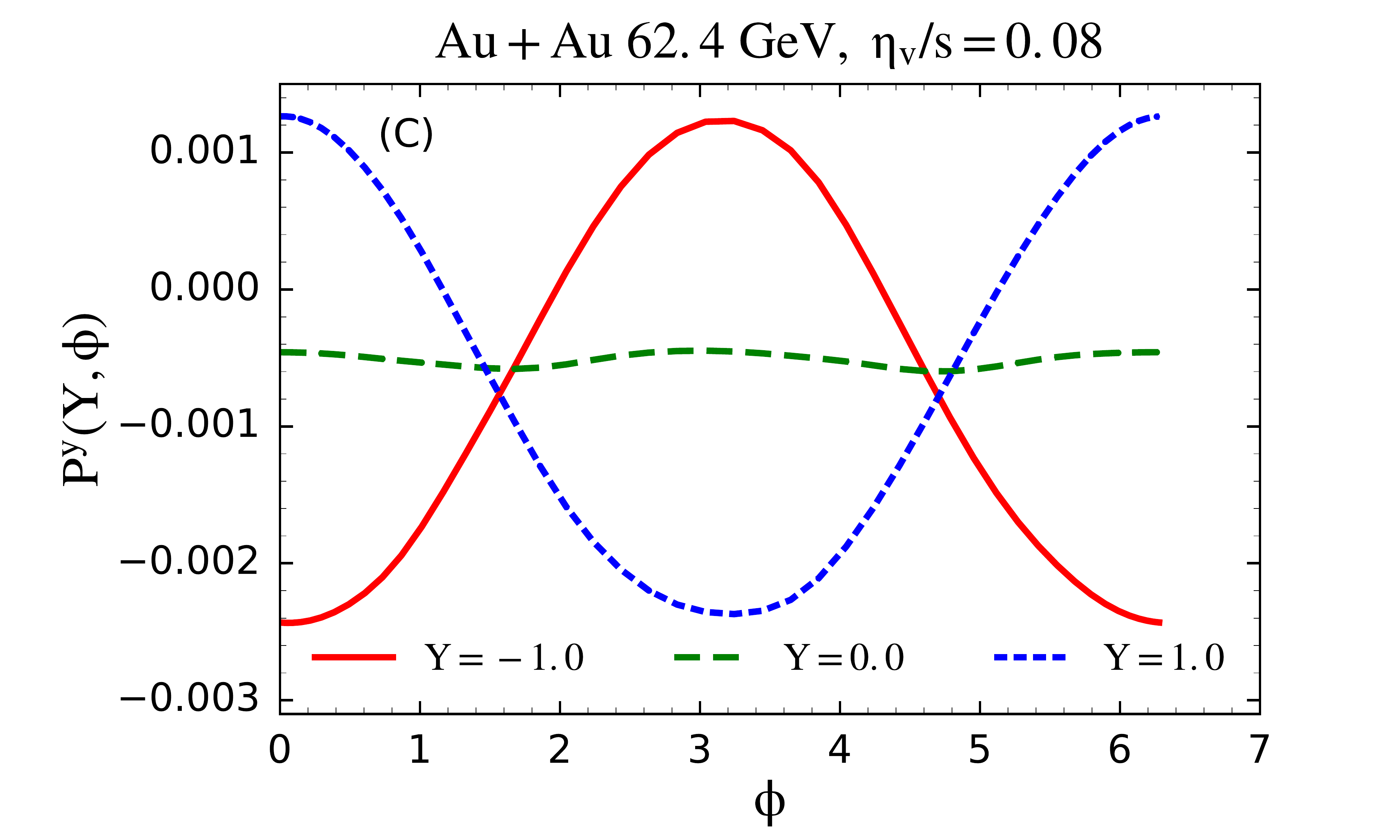}
\caption{The rapidity and azimuthal angle distribution of $\Lambda$ spin projected to $y$ direction for Pb+Pb $\sqrt{s_{NN}}=2.76$~TeV (left), Au+Au $\sqrt{s_{NN}}=200$~GeV (middle) and Au+Au $\sqrt{s_{NN}}=62.4$~GeV (right) collisions. }
\label{fig:pol_vs_phi}
\end{center}
\end{figure}
 As shown in Fig.~\ref{fig:pol_vs_phi}, the polarization are shifted to the $-y$ direction which is the direction of the global angular momentum. Locally, the azimuthal angle distribution for $P^{y}$ has a cosine structure which indicates a vortex ring~\cite{Pang:2016igs}. The helicity of the vortex ring is opposite for forward and backward rapidity. The polarization at mid-rapidity shows a weaker azimuthal angle dependence. The maximal magnitude of the polarization at $62.4$~GeV is about 5 times that at $2.76$~TeV.

\begin{figure}[htbp]
\begin{center}
\includegraphics[width=0.32\textwidth, trim={1.5cm 0.55cm 3cm 0.6cm},clip]{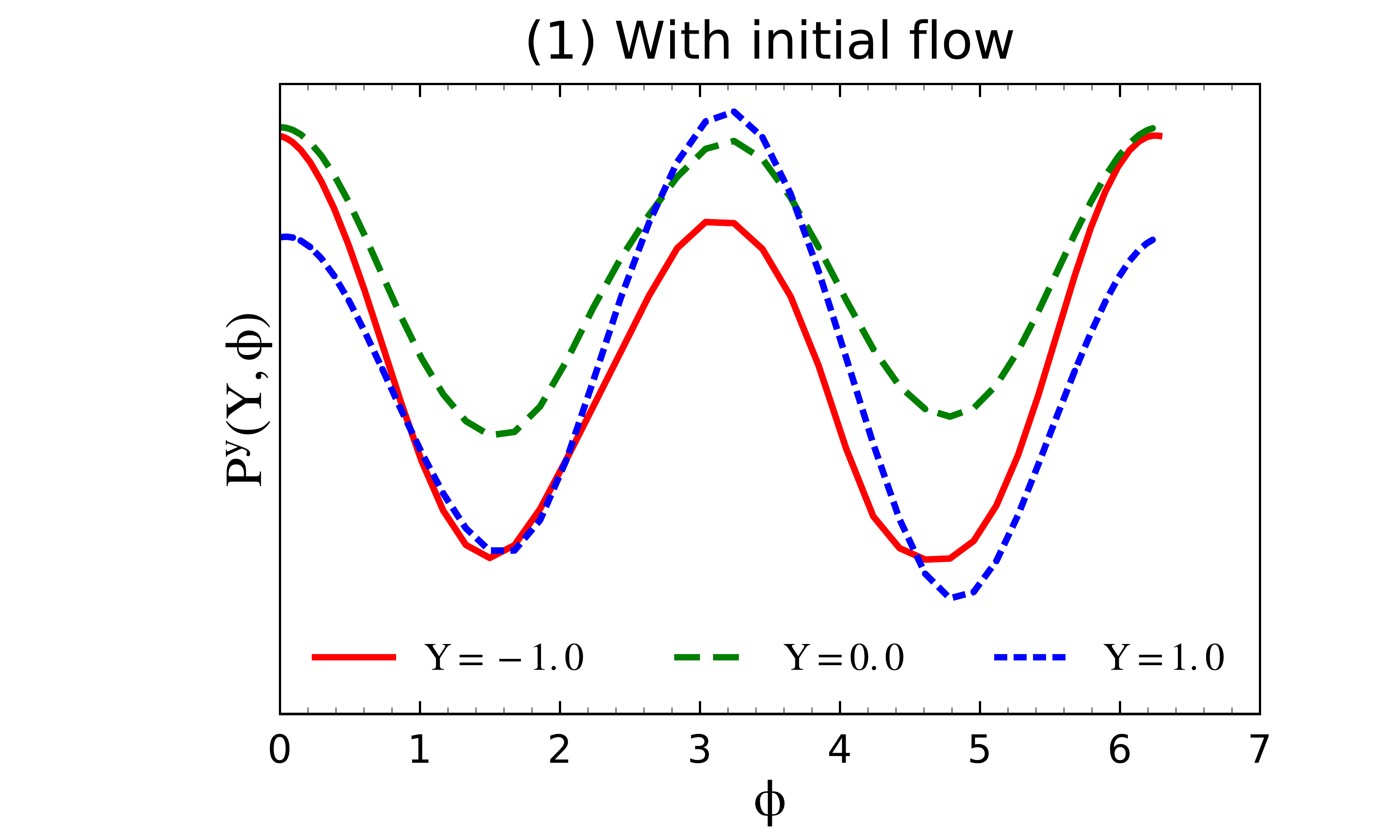}\hspace{0.1\textwidth}\includegraphics[width=0.32\textwidth, trim={1.5cm 0.55cm 3cm 0.6cm},clip]{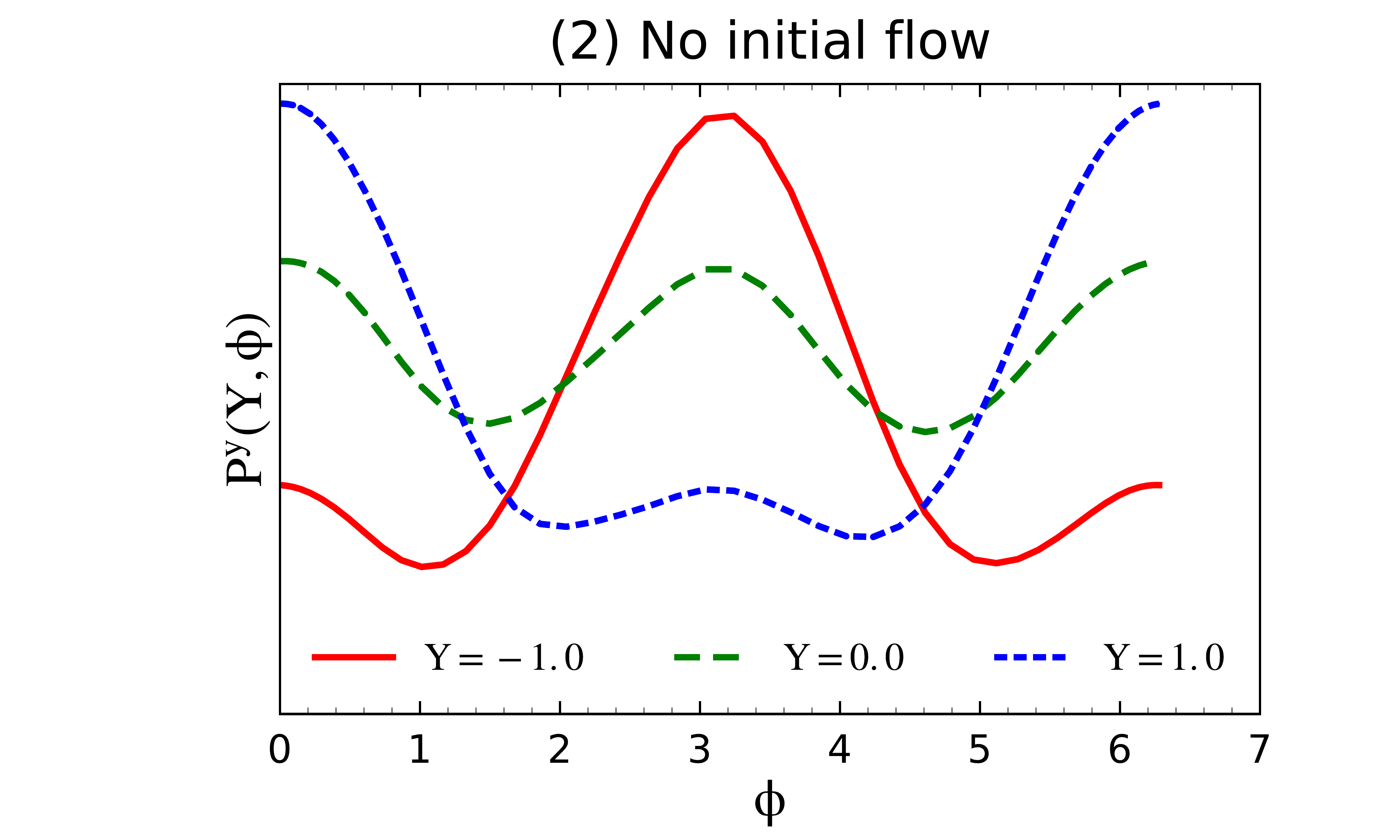}
\caption{The rapidity and azimuthal angle distribution of $\Lambda$ spin projected to $y$ direction $P^{y}$ for (1) with initial transverse flow, (2) without initial transverse flow.
}
\label{fig:pol_ini_flow}
\end{center}
\end{figure}

For comparison, we compute the local polarization of $\Lambda$ by the CLVisc model with the initial transverse flow $v_x$ and $v_y$ given by the energy-momentum tensor $T^{\mu\nu}$ of initial partons by the AMPT. With this configuration, the deposited initial angular momenta are not only given by the matter asymmetry (between $x>0$ and $x<0$) but also by the $v_x$ gradients along $\eta$. As a result, the local polarization at $\Phi=0$ now has the similar magnitude to that at $\Phi=\pi$ as shown in Fig.~\ref{fig:pol_ini_flow}. This dramatic change indicates that the local polarization of $\Lambda$ hyperons around mid-rapidity may provide rigorous constraints on the initial transverse flow.

\section{Global $\Lambda$ polarization from AMPT}
We also compute the energy dependence of the global $\Lambda$ polarization within the AMPT model. The fluid velocity and vorticity field are computed from the event average of the four-momentum of all particles in each space-time cell. We consider two impact parameters $b=7$~fm and 9~fm at each specific collision energy. The product of the vorticity field $\Omega_{zx}$ and the $\Lambda$ distribution $f_{\Lambda}$ is integrated over on the hyper-surface to obtain the global $\Lambda$ polarization. The numerical result with both primary $\Lambda$ and feed-down contributions agrees with experimental data semi-quantitatively. The energy dependence is investigated in details using the distribution of $f_{\Lambda}$ and $\Omega_{zx}$. A visualization of these two quantities in the reaction plane indicates that the angular momentum deposition at mid-rapidity is quite small for high beam energies, which is consistent with the Bjorken scaling scenario. This scaling is broken for low beam energies with large asymmetry between the forward and backward going participants, which gives rise to a tilted shape at the mid-rapidity, see Ref. \cite{Li:2017slc} for details.

\begin{figure}[htbp]
\begin{center}
\includegraphics[width=0.32\textwidth, trim={1.5cm 0.5cm 2cm 1.5cm},clip]{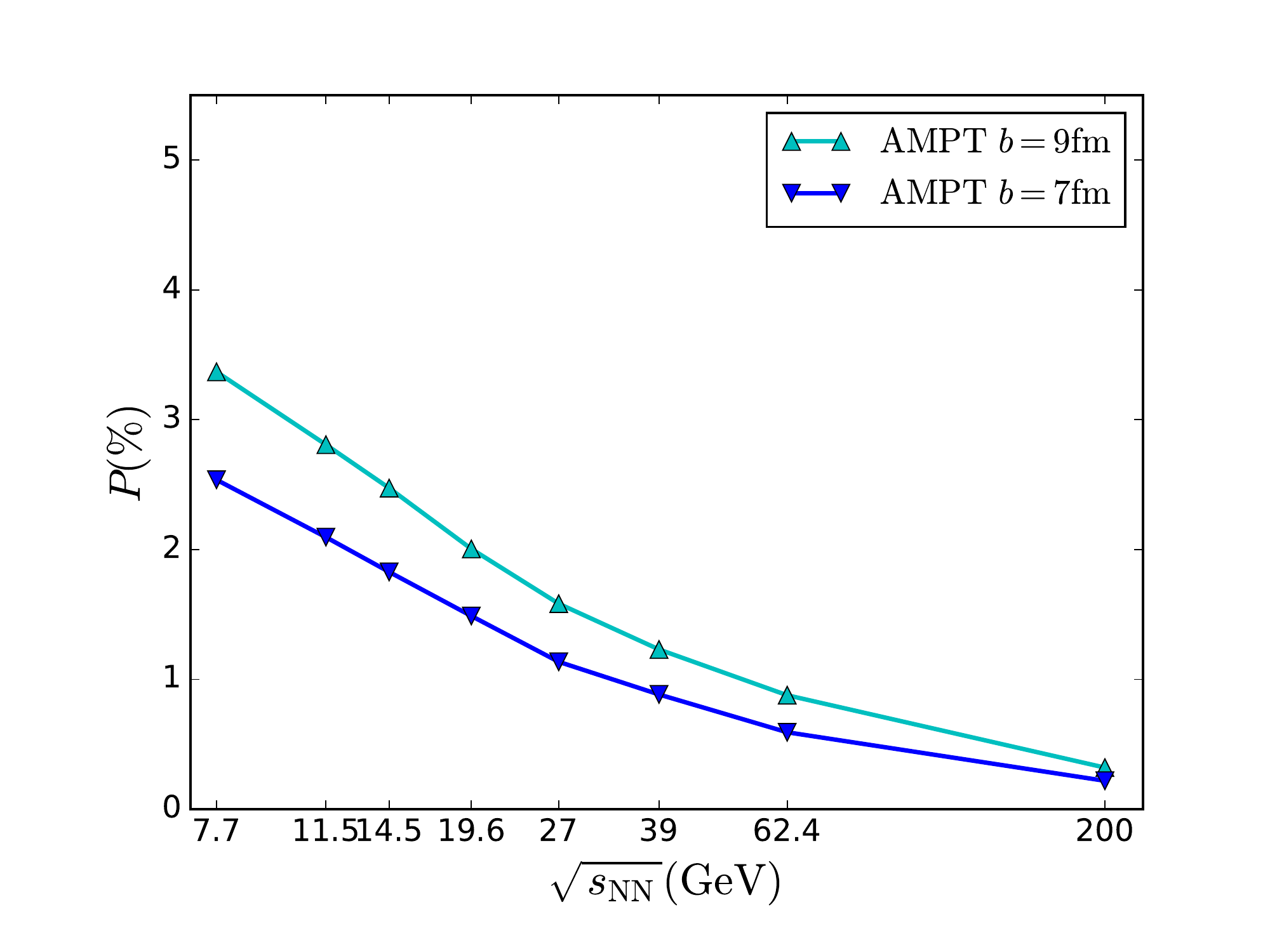}\hspace{0.1\textwidth}\includegraphics[width=0.32\textwidth, trim={1.5cm 0.5cm 2cm 1.5cm},clip]{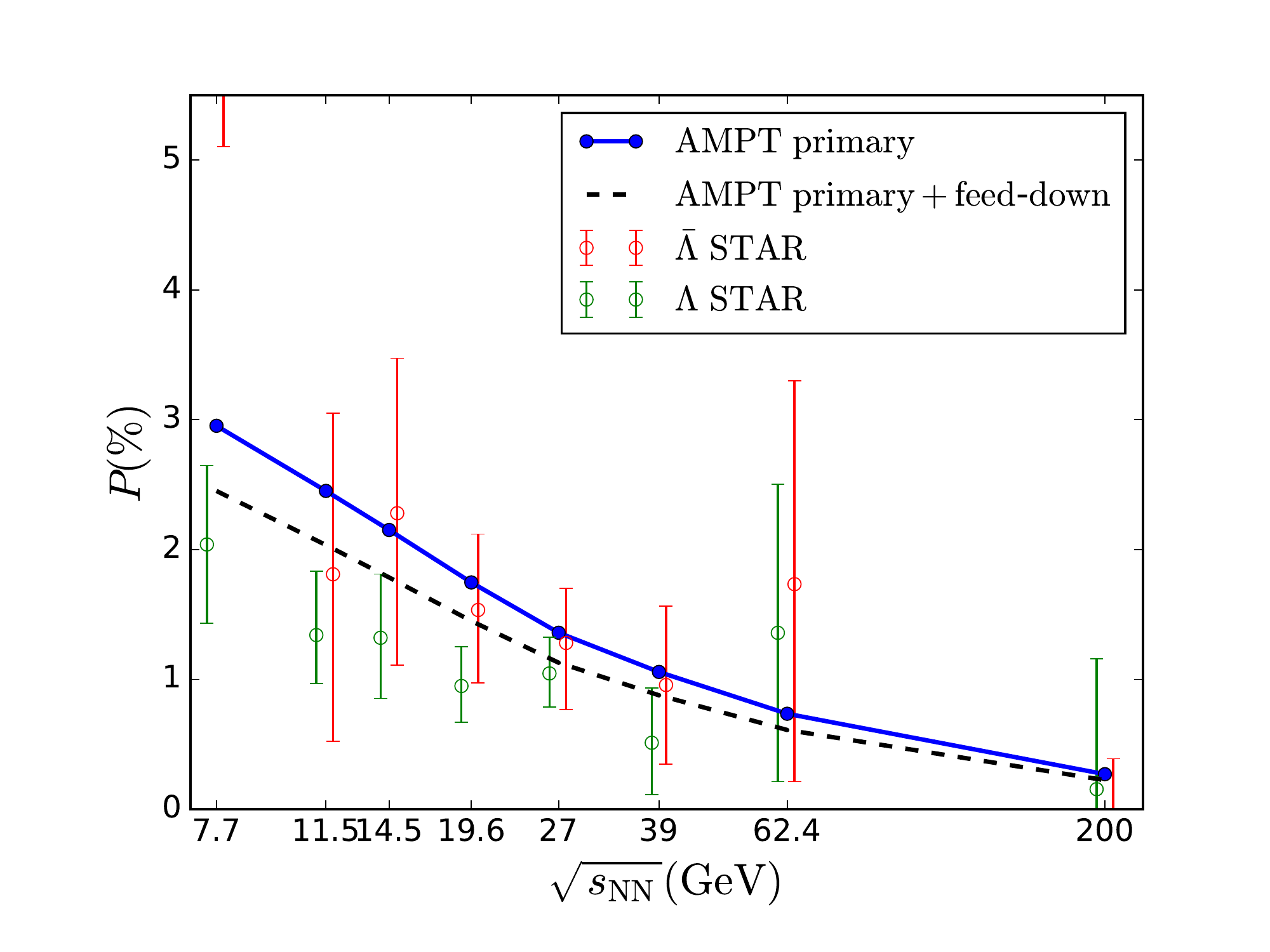}
\caption{(Left) The global polarization of $\Lambda$  from AMPT model for 2 different impact parameters $b=7$~fm and 9~fm (approximating $20-50\%$ collisions). (Right) The model data comparison for with and without feed-down contribution from $\Sigma$ decay.}
\label{fig:global_polarization_ampt}
\end{center}
\end{figure}

\section{Acknowledgements}
HL, QW and XLX are supported by the MSBRD in China under the Grant No. 2015CB856902 and 2014CB845402 and by the NSFC under the Grant No.11535012.  LGP and HP acknowledge funding of a Helmholtz Young Investigator Group VH-NG-822 from the Helmholtz Association and GSI, the Helmholtz International Center for the Facility for Antiproton and Ion Research (HIC for FAIR) within the
framework of the Landes-Offensive zur Entwicklung Wissenschaftlich-Oekonomischer Exzellenz (LOEWE) program launched by the State of Hesse. XNW is supported by NSFC under the Grant Nos. 11221504 and 11535012, by MOST of China under Grant No.2014DFG02050, by the MSBRD under the Grant No. 2015CB856902, 2014CB845404, and 2014CB845406, by U.S. DOE under Contract No. DE-AC02-05CH11231.

\bibliographystyle{elsarticle-num}
\bibliography{inspire}


\end{document}